\documentclass{moriond}

\newcommand{\Photo}{}%\includegraphics[height=35mm]{photo}}

\newcommand{\elp}{{e^+}}
\newcommand{\elm}{{e^-}}
\newcommand{\mup}{{\mu^+}}
\newcommand{\mum}{{\mu^-}}
\newcommand{\kp}{{K^+}}

\newcommand{\pip}{{\pi^+}}
\newcommand{\pim}{{\pi^-}}
\newcommand{\jp}{{J/\psi}}
\newcommand{\psp}{{\psi'}}
\newcommand{\ups}{{\Upsilon}}
\newcommand{\z}[1]{{Z_c(#1)^+}}
\newcommand{\br}{\mathcal{B}}
\newcommand{\gev}{\mathrm{GeV}}
\newcommand{\gevcc}{\mathrm{GeV}/c^2}
\newcommand{\mev}{\mathrm{MeV}}
\newcommand{\mevcc}{\mathrm{MeV}/c^2}

\begin{document}
\vspace*{4cm}
\title{RECENT RESULTS ON QUARKONIUM-(LIKE) STATES AT BELLE}

\author{K. CHILIKIN (FOR THE BELLE COLLABORATION)}

\address{Institute for Theoretical and Experimental Physics, \\
117218, Bolshaya Cheremuskinskaya 25, Moscow, Russia}

\maketitle\abstracts{
Recent results on quarkonium and quarkonuim-like states at Belle are presented.
}

\section{Bottomonium results}

\subsection{Measurement of $\elp \elm \to b \bar{b}$ and
$\elp \elm \to \ups \pip \pim$ crossections}

\begin{figure}[h]
\begin{center}
\includegraphics[width=6cm]{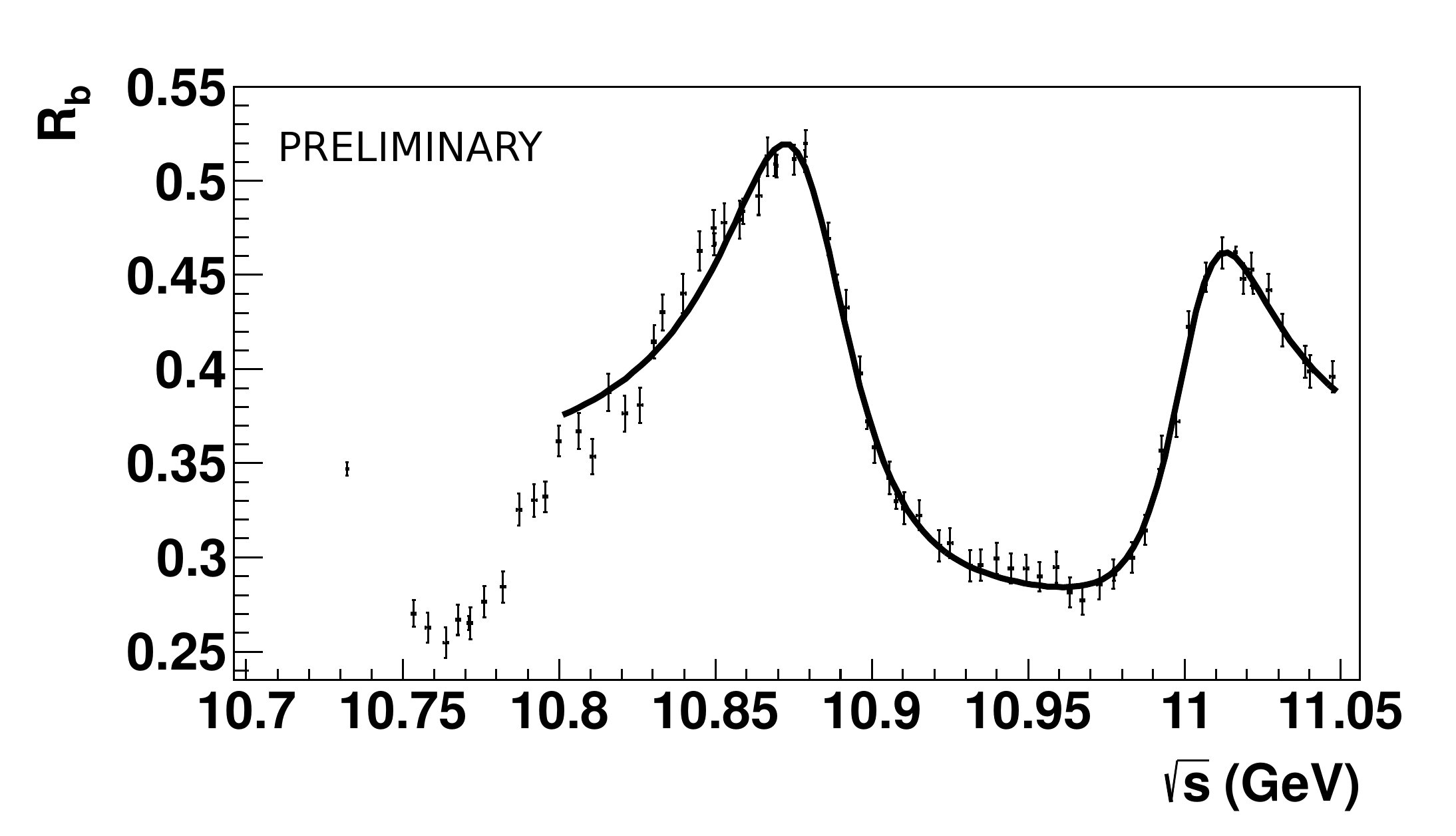}
\includegraphics[width=6cm]{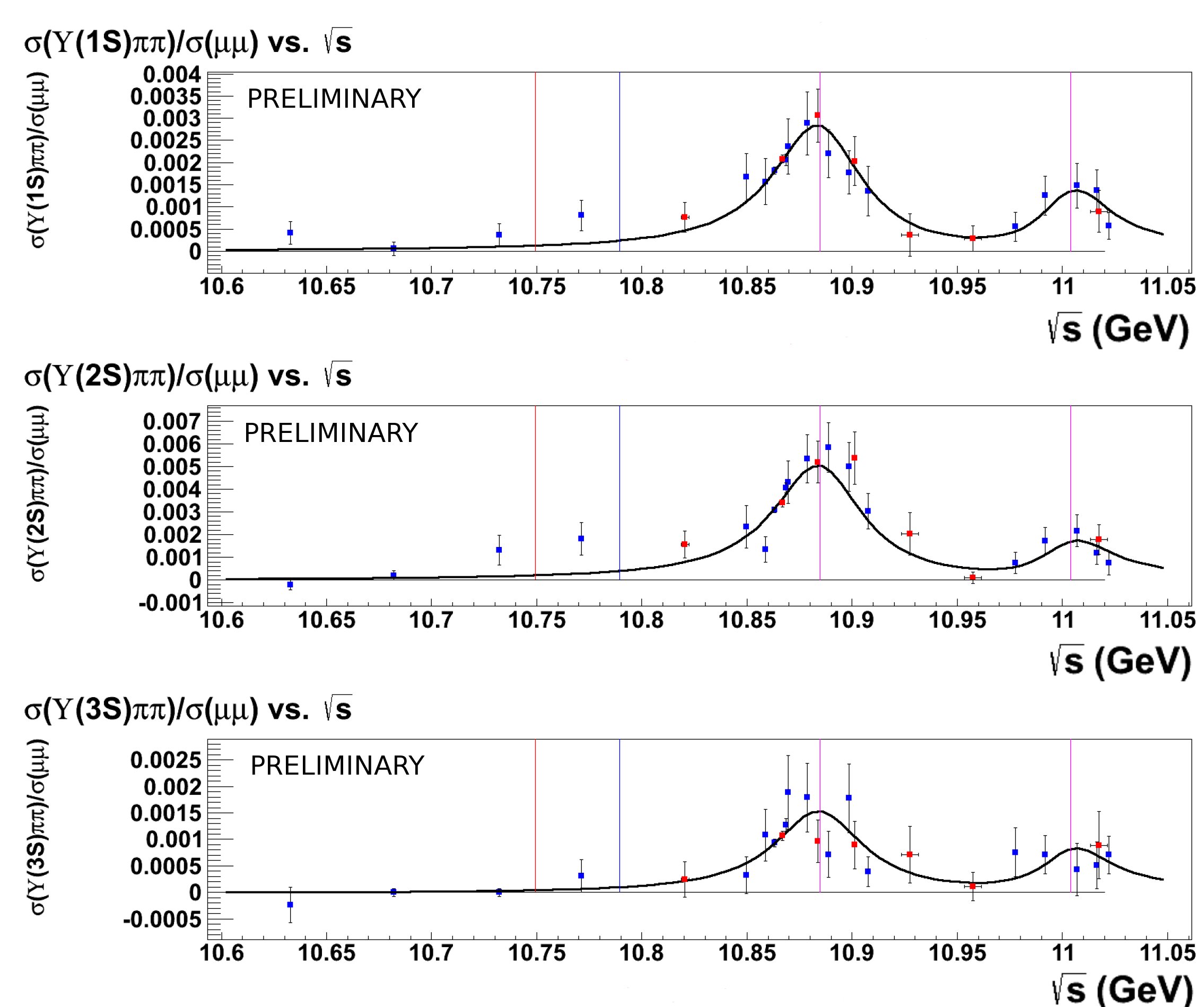}
\end{center}
\caption{Crossections of $\elp \elm \to b \bar{b}$ and $\elp \elm \to \ups \pip \pim$.}
\label{fig:ypspipi}
\end{figure}

The crossections of $\elp \elm \to b \bar{b}$ and $\elp \elm \to \ups \pip \pim$
in the region of the $\ups(5S)$ and $\ups(6S)$ resonances were measured.
The results are shown in Fig.~\ref{fig:ypspipi}. Decays of the $\ups(6S)$
to $\ups(1S,2S,3S) \pip \pim$ are observed.
The $R_X$ (ratios of the
process $X$ crossection to $\elp \elm \to \mup \mum$ crossection) at
$10.865\ \gev$ are found to be:
$R_{\ups(1S) \pip \pim} = (2.4 \pm 0.1 \pm 0.7) \times 10^{-3}$,
$R_{\ups(2S) \pip \pim} = (4.2 \pm 0.2 \pm 1.1) \times 10^{-3}$,
$R_{\ups(3S) \pip \pim} = (1.3 \pm 0.1 \pm 0.5) \times 10^{-3}$.
The mass of the $Y(5S)$ is measured; the result is
$10880.4 \pm 0.9 \pm 1.4\ \mevcc$ for the $b \bar{b}$ sample and
$10884.6 \pm 1.4 \pm 1.1\ \mevcc$ for the $\ups \pip \pim$ sample.

\subsection{Observation of $\ups(4S) \to h_b \eta$}

\begin{figure}[h]
\begin{center}
\includegraphics[width=6cm]{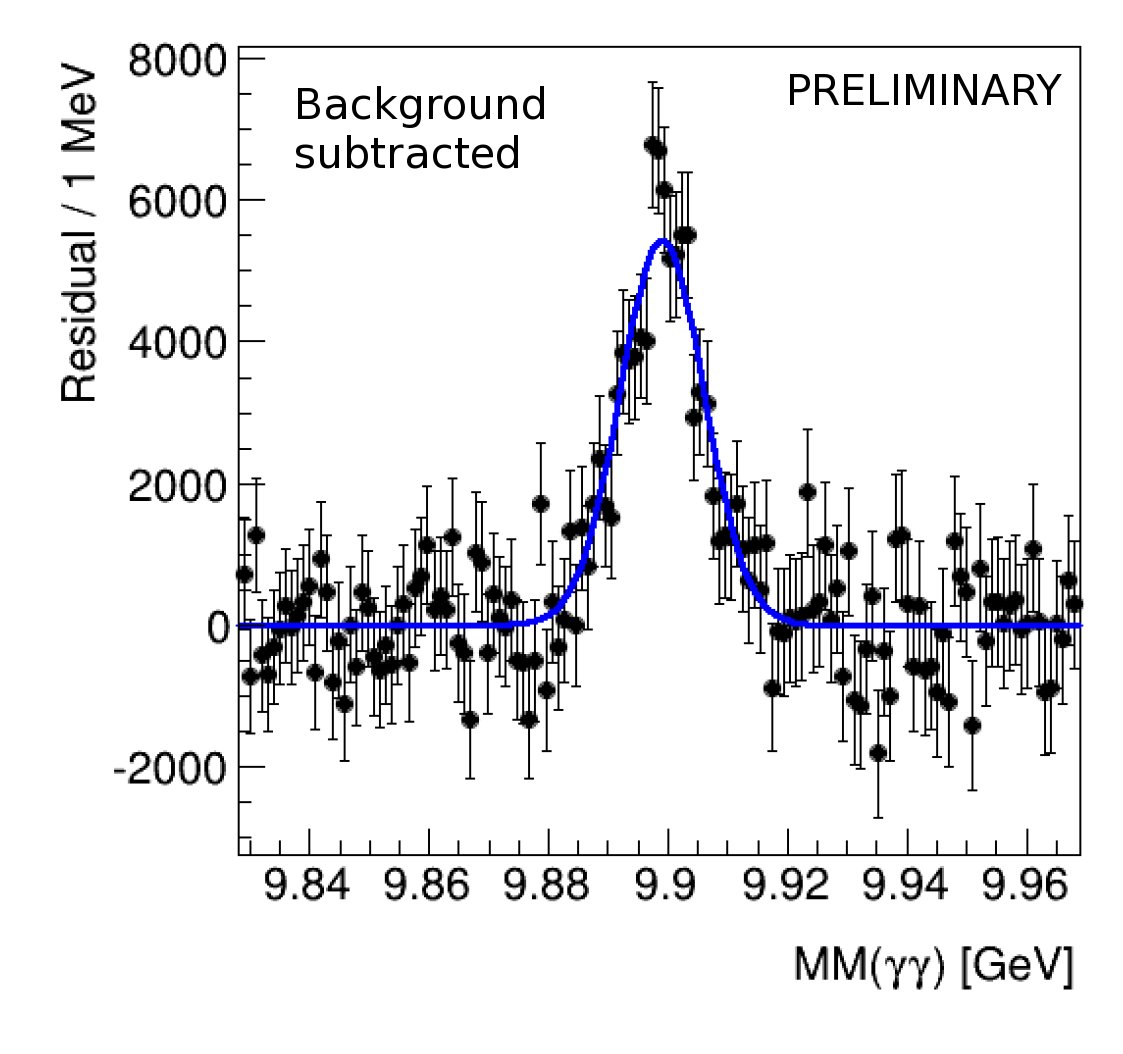}
\includegraphics[width=6cm]{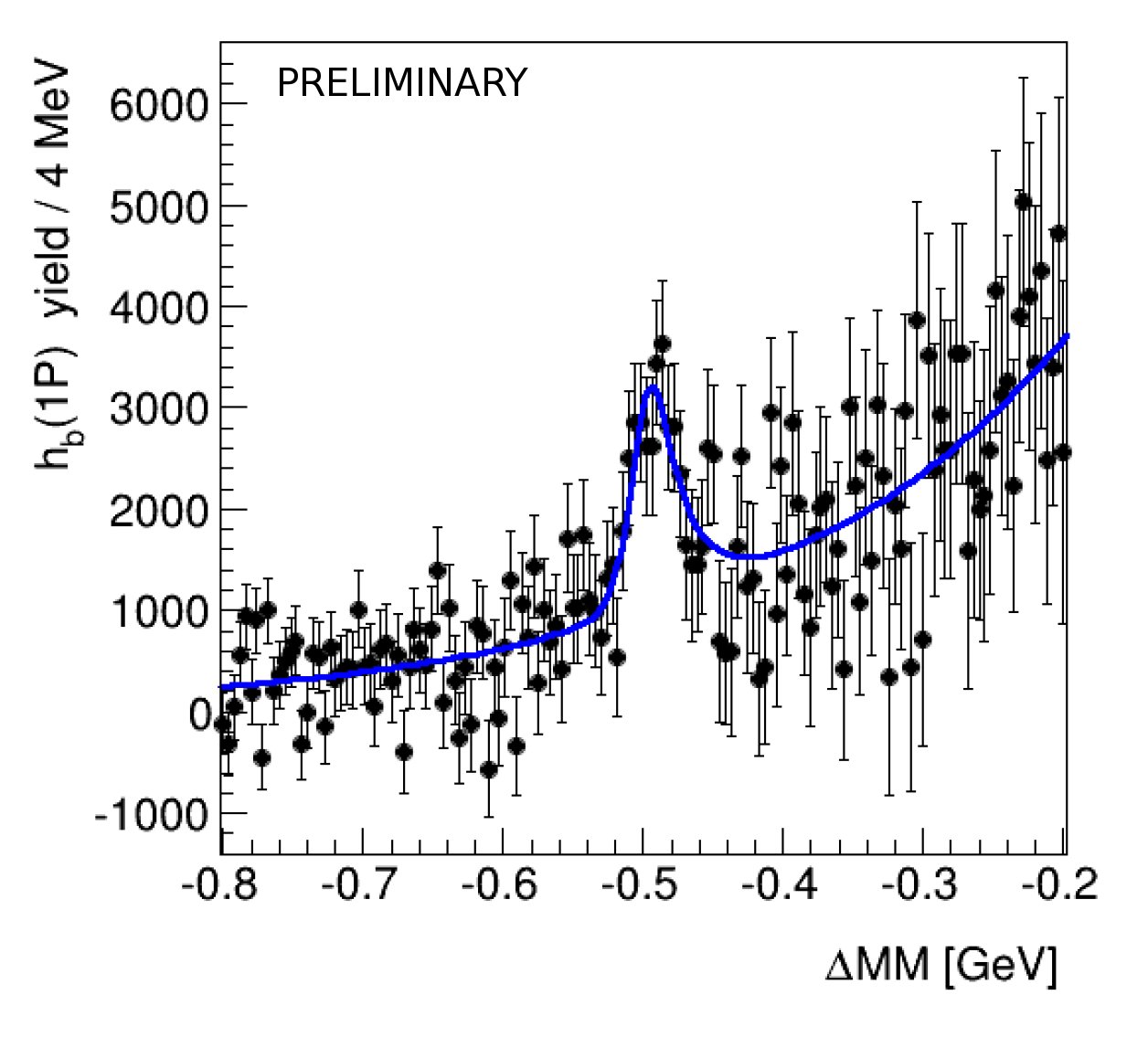}
\end{center}
\caption{The $\eta$ recoil mass and the difference of $\eta \gamma$ and $\eta$ recoil masses.}
\label{fig:etahb}
\end{figure}

The decay $\ups(4S) \to h_b \eta$ was observed. The $\eta$ is reconstructed
in $\eta \to \gamma \gamma$ decay mode.
The background-subtracted
distribution of the $\eta$ recoil mass is shown in Fig.~\ref{fig:etahb};
the signal significance is $17\sigma$.
The branching fraction is found to be
$\br(\ups(4S) \to h_b \eta) = (1.83 \pm 0.16 \pm 0.17) \times 10^{-3}$.
The same process with $h_b \to \gamma \eta_b$ is also studied. The
difference of $\eta \gamma$ and $\eta$ recoil masses is shown in
Fig.~\ref{fig:etahb}. The branching fraction of $h_b \to \gamma \eta_b$ is
found to be $(52.8\pm4.7\pm4.2)\%$; the difference of $\eta_b$ and $h_b$
masses is $M(\eta_b) - M(h_b) = -494.0\pm1.3\pm2.8\ \mevcc$, and
$\eta_b$ width is $\Gamma(\eta_b) = 10.7\pm2.3\pm3.7\ \mev$.

\section{Charmonium results}

\subsection{Observation of $\z{4430}$}

\begin{figure}[h]
\begin{center}
\includegraphics[width=6cm]{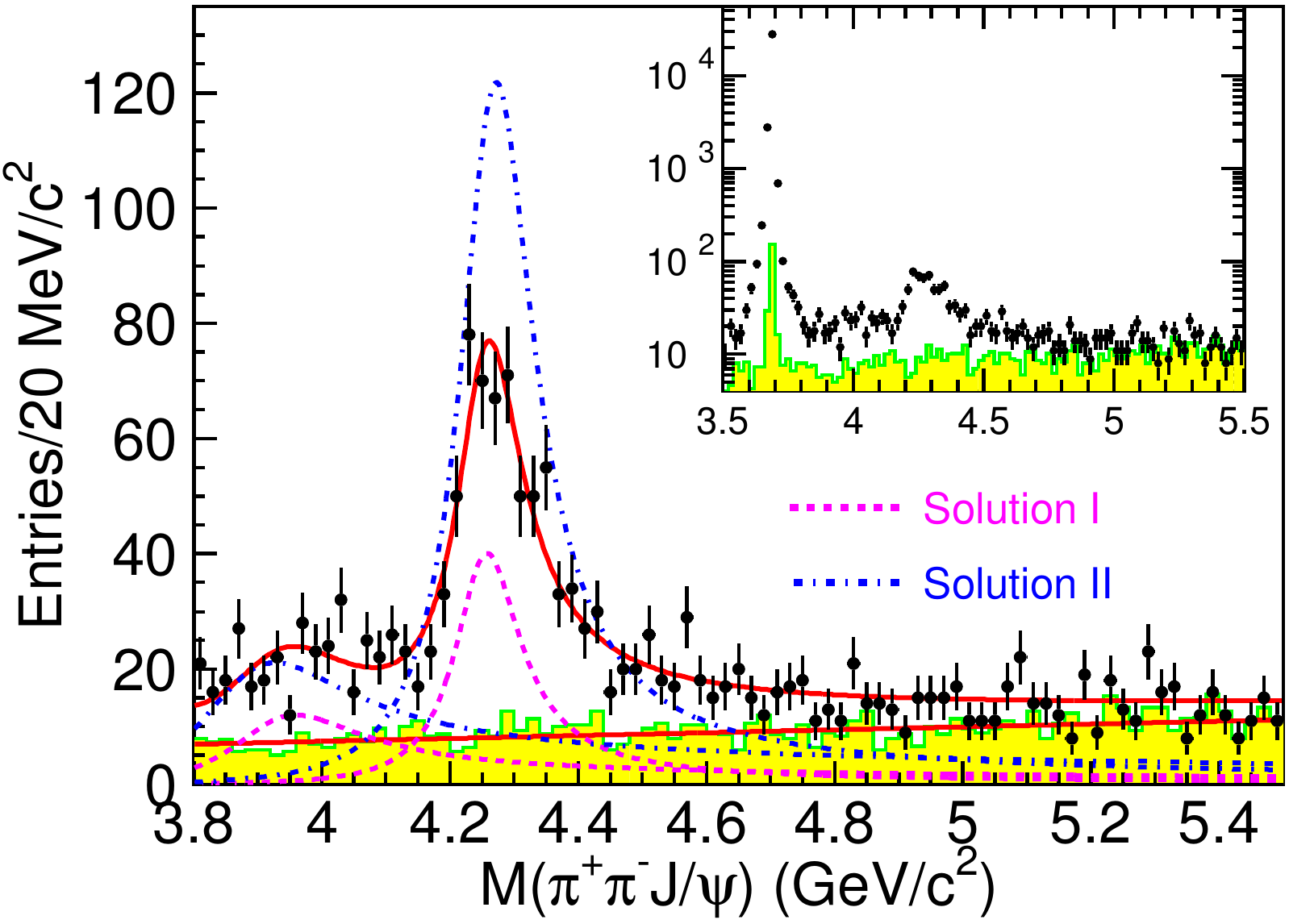}
\includegraphics[width=6cm]{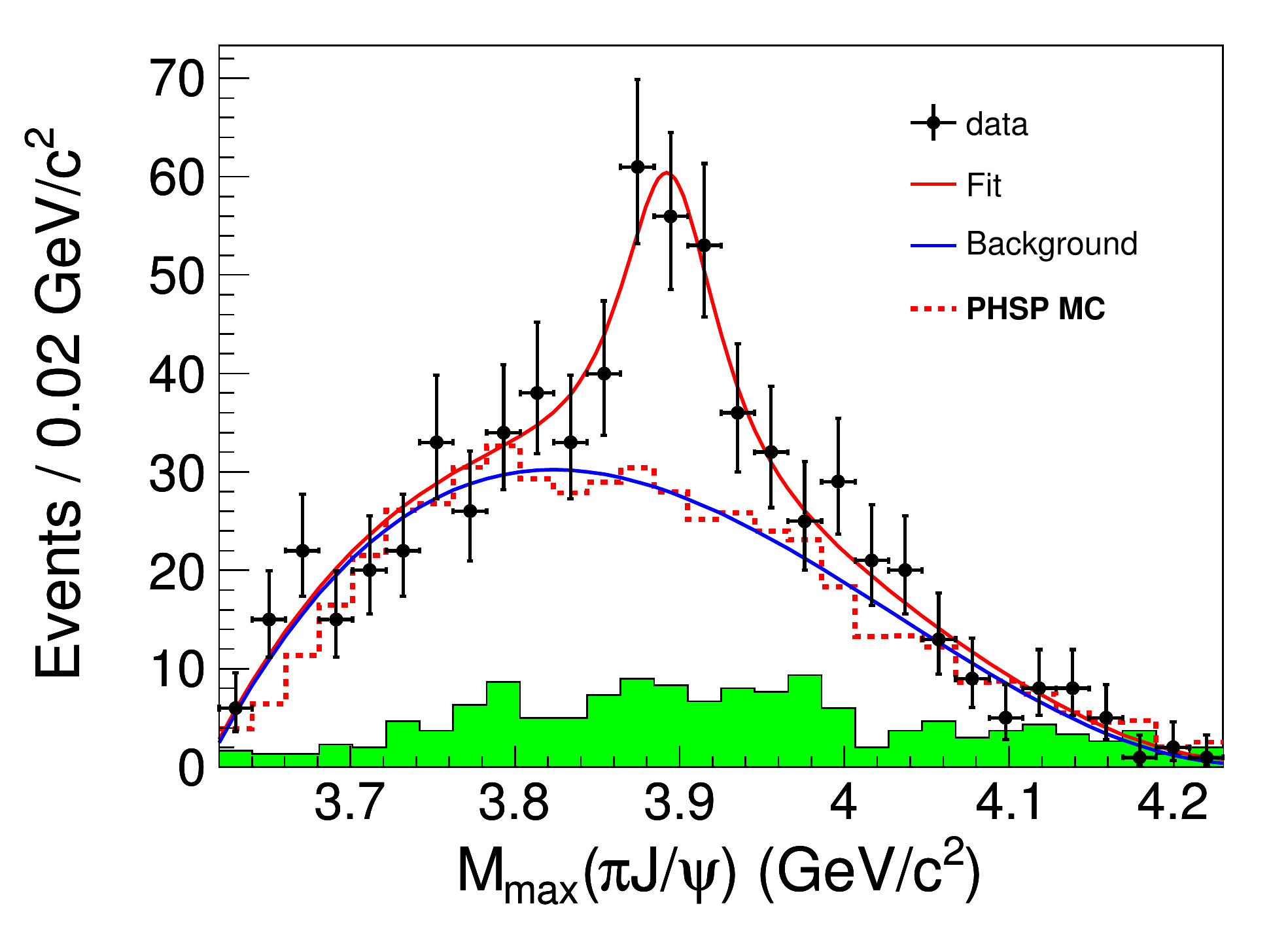}
\end{center}
\caption{Crossection of $\jp \pi^+ \pi^-$ via ISR and maximal $\jp \pi$
invariant mass.}
\label{fig:z3900}
\end{figure}

The $\z{3900}$ was observed in $\jp \pi$ invariant mass distribution
in $Y(4260) \to \jp \pip \pim$~\cite{z3900belle} (this state was also observed
by BESIII Collaboration~\cite{z3900bes} simultaneously).
The crossection of $\jp \pi^+ \pi^-$ via ISR
and the distribution of the maximal $\jp \pi$
invariant mass ($\max(M(\jp \pip), M(\jp \pim))$) in the $Y(4260)$ region
defined as $4.15\ \gevcc < M(\jp \pip \pim) < 4.45\ \gevcc$ are shown
in Fig.~\ref{fig:z3900}. The resulting parameters of the $\z{3900}$ are
$M = 3894.5 \pm 6.6 \pm 4.5\ \mev$, $\Gamma = 63 \pm 24 \pm 26\ \mev$,
$\br(Y(4260)\to Z_c(3900)^\pm\pi^\pm)\times\br(Z_c(3900)^\pm\to\pi^\pm\jp)/
\br(Y(4260)\to\jp\pip\pim) = \\(29.0 \pm 8.9)\%$.

\subsection{Analysis of $B^\pm \to \chi_{c1} \pip \pim K^\pm$}

\begin{figure}[h]
\begin{center}
\includegraphics[width=6cm]{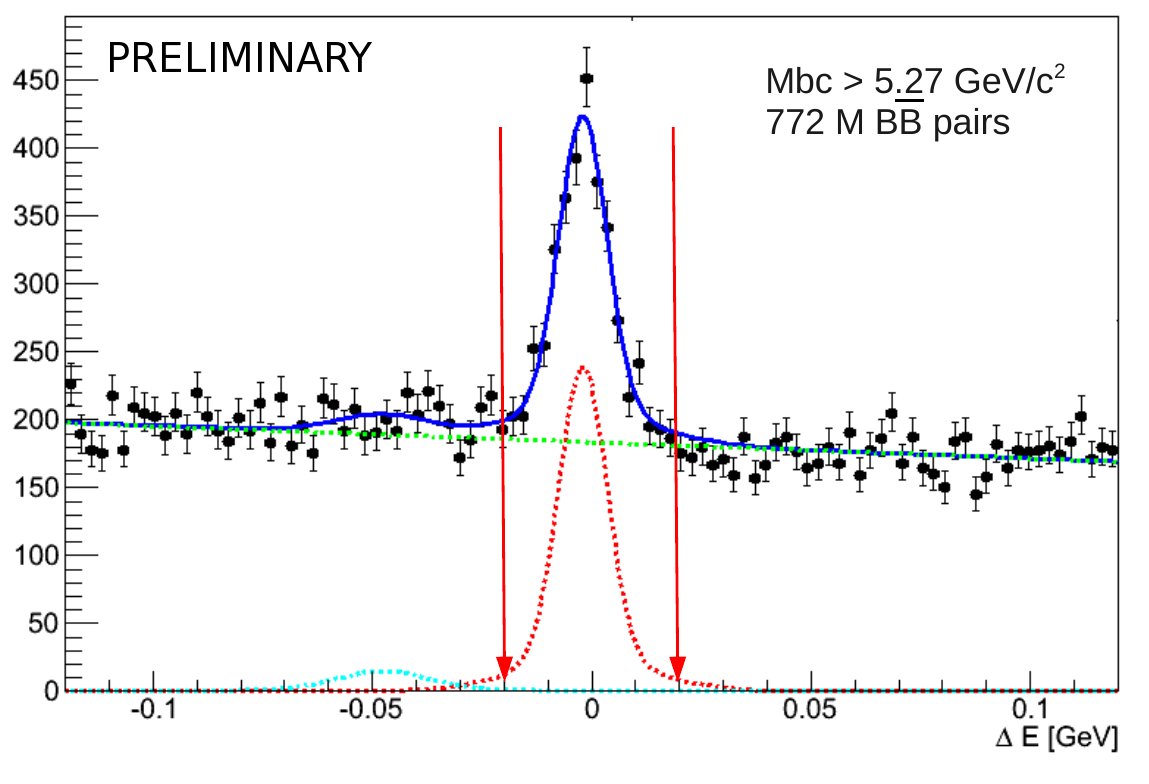}
\includegraphics[width=6cm]{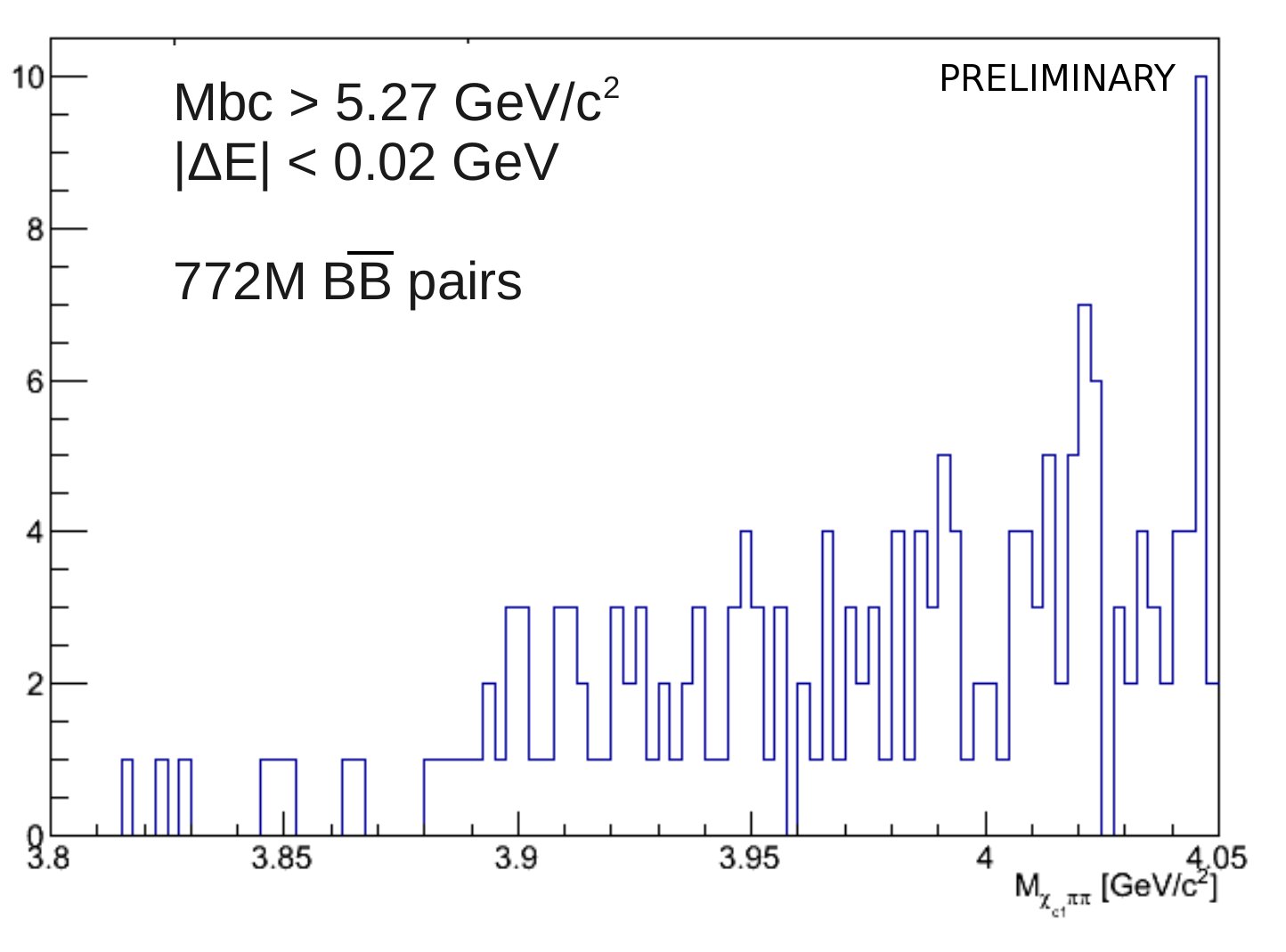}
\end{center}
\caption{The $\Delta E$ distribution and the $\chi_{c1} \pip \pim$ invariant
mass distribution in the signal region.}
\label{fig:chicpipi}
\end{figure}

The decay mode $B^\pm \to \chi_{c1} \pip \pim K^\pm$ was observed (the
significance is about $20\sigma$; the $\Delta E$ distribution is shown
in Fig.~\ref{fig:chicpipi}). The branching fraction is
$\br(B\to\chi_{c1}\pip\pim K) = (3.94\pm0.19\pm0.30)\times10^{-4}$.
A search
for resonances decaying to $\chi_{c1} \pip \pim$ was performed. The distribution
of $\chi_{c1} \pip \pim$ invariant mass is shown in Fig.~\ref{fig:chicpipi}.
No significant signal is found.

\subsection{Observation of $B \to X(3872) K \pi$}

\begin{figure}[h]
\begin{center}
\includegraphics[width=5cm]{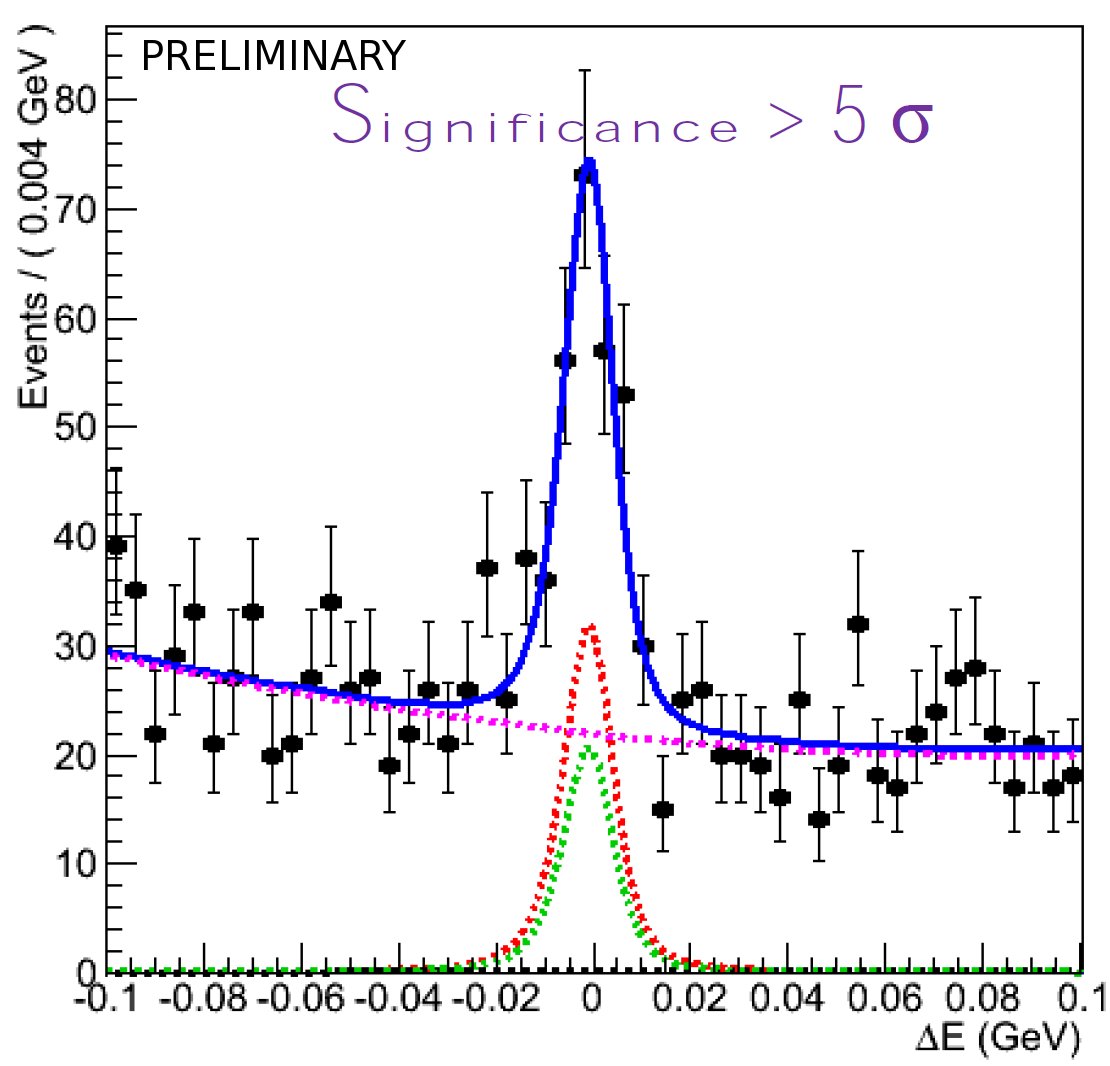}
\includegraphics[width=5cm]{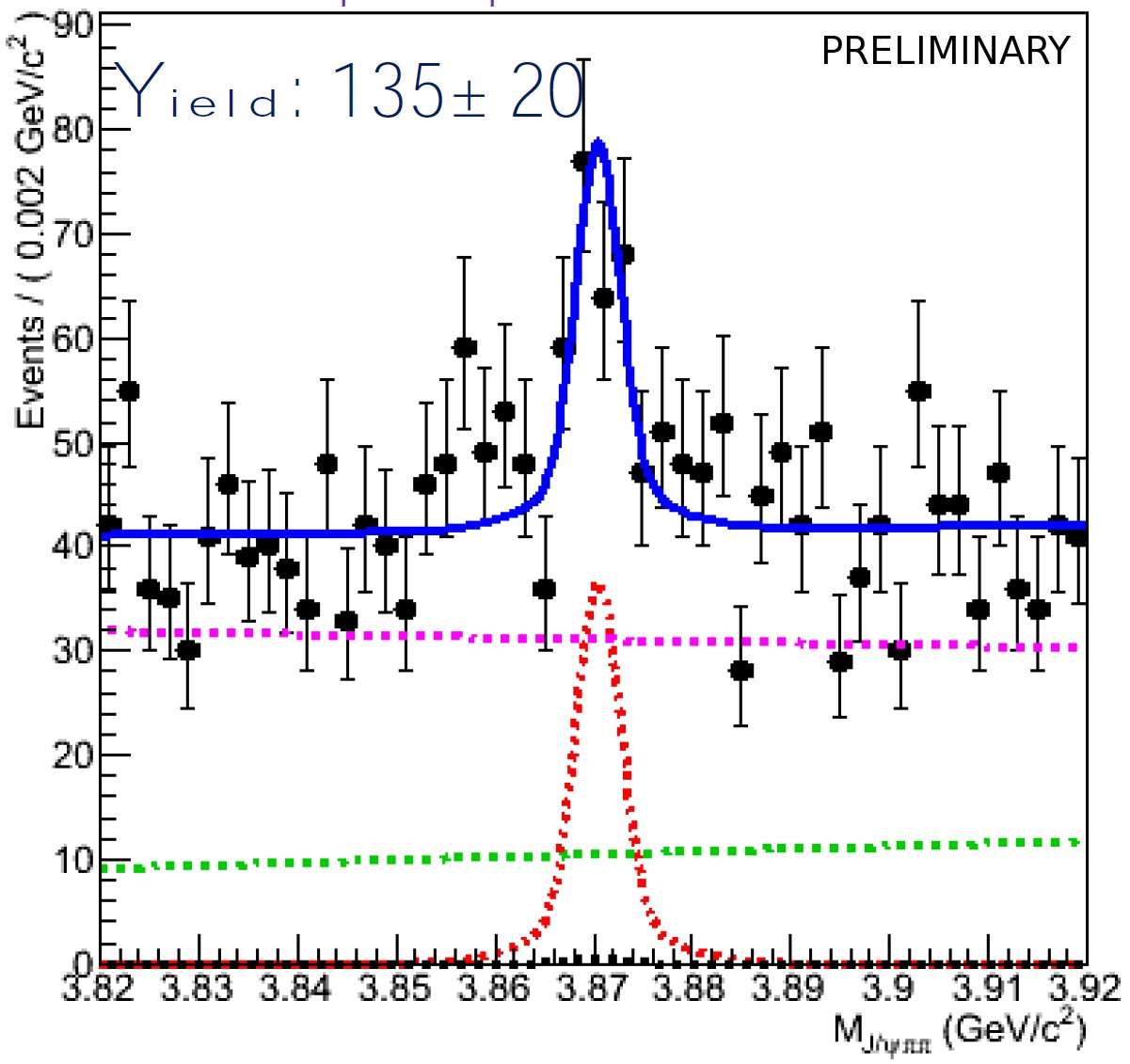}
\end{center}
\caption{The $\Delta E$ distribution and the $\jp \pip \pim$ invariant mass
distribution in the signal region.}
\label{fig:x3872}
\end{figure}

The decay mode $B^0 \to X(3872) \kp \pim$ was observed. The $\Delta E$
distribution and the $\jp \pip \pim$
invariant mass distribution are shown in Fig.~\ref{fig:x3872}.
The branching fraction product is found to be
$\br(B^0 \to X(3872)\kp\pim)\times\br(X(3872)\to\jp\pip\pim)=
(8.6\pm1.3^{+0.5}_{-0.8})\times10^{-6}$. The $K^*(892)$ fraction is
$0.29\pm0.08$.

\subsection{Measurement of the $\z{4430}$ quantum numbers}

\begin{figure}[h]
\begin{center}
\includegraphics[width=6cm]{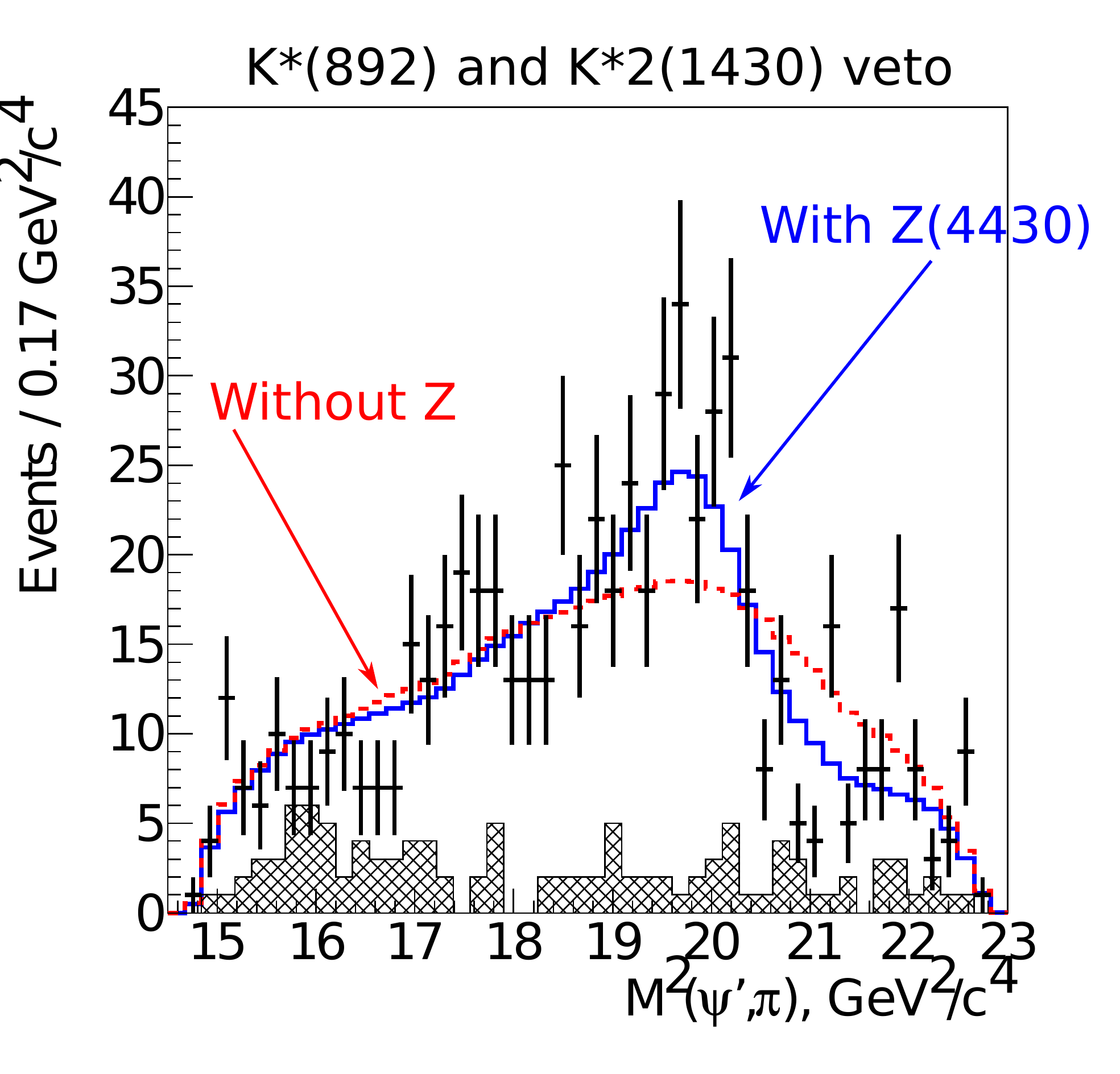}
\end{center}
\caption{Projection of the fit results with and without the $\z{4430}$.}
\label{fig:z4430}
\end{figure}

An amplitude analysis of $B^0 \to \psp \kp \pim$ was performed,
and the spin and parity of the $\z{4430}$ has been measured
~\cite{z4430jp}. The $\z{4430}$
signal is added to the amplitude with different $J^P$ hypotheses. The
preferred assignment of the quantum numbers is $1^+$.
The exclusion levels of the $0^-$, $1^-$, $2^-$ and $2^+$ hypotheses
are $3.4\sigma$, $3.7\sigma$, $4.7\sigma$ and $5.1\sigma$, respectively.
Projection of the fit results with and without the $\z{4430}$ ($J^P=1^+$)
onto $M^2(\psp \pi)$ is shown in Fig.~\ref{fig:z4430}. The mass and width
of the $\z{4430}$ are $M = 4485^{+22 +28}_{-22 -11}\ \mevcc$,
$\Gamma = 200^{+41 +26}_{-46 -35}\ \mev$.

\subsection{Amplitude analysis of $B \to \jp K \pi$}

\begin{figure}[h]
\begin{center}
\includegraphics[width=6cm]{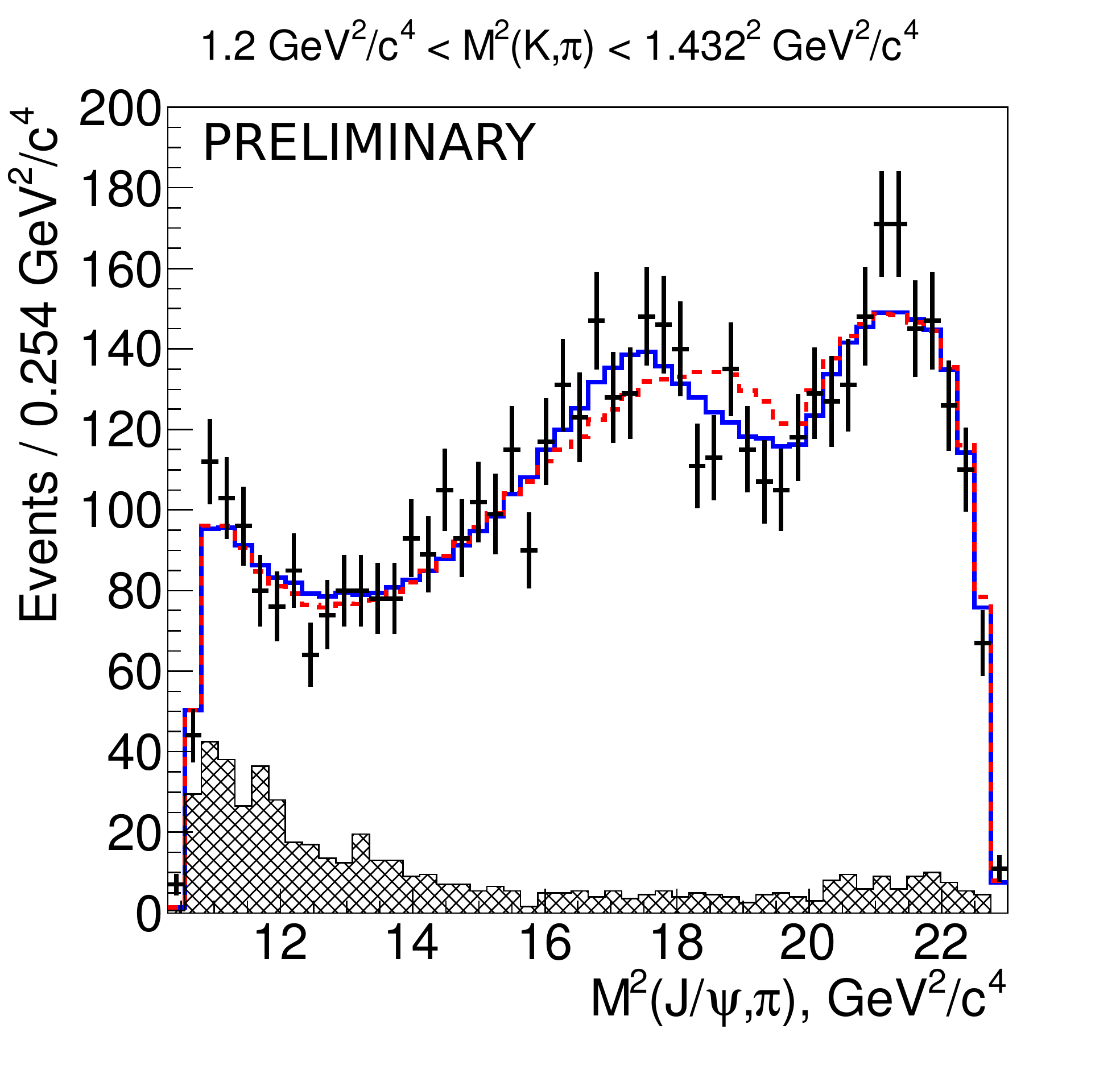}
\includegraphics[width=6cm]{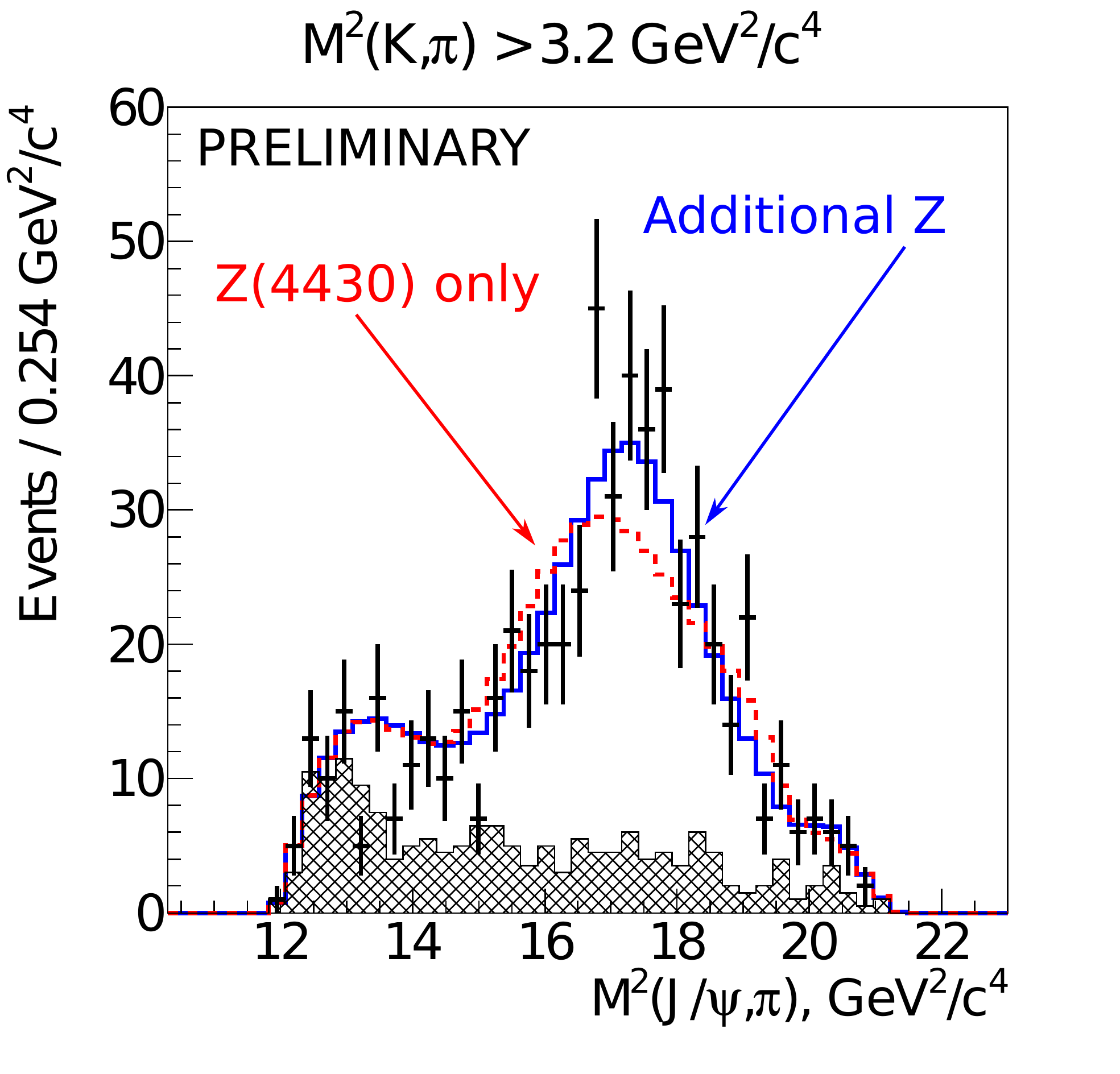}
\end{center}
\caption{Projection of the fit results with and without the $\z{4200}$.}
\label{fig:jpkpi}
\end{figure}

An amplitude analysis of $B^0 \to \jp \kp \pim$ was performed. Projections
of the fit results onto $M^2(\jp \pi)$ are shown in Fig.~\ref{fig:jpkpi}.
A new $Z_c^+$ state is observed with $7.2\sigma$ significance;
the preferred quantum numbers are $J^P=1^+$. It is referred to as the
$\z{4200}$.
The exclusion levels of the $0^-$, $1^-$, $2^-$, $2^+$ hypotheses are
$6.7\sigma$, $7.7\sigma$, $5.2\sigma$, $7.6\sigma$. The mass and width
of the $\z{4200}$ are $M = 4196^{+31 +17}_{-29 -6}\ \mevcc$,
$\Gamma = 370^{+70 +70}_{-70 -85}\ \mev$. In addition, evidence for
$\z{4430}\to \jp \pip$ is found.

\section*{Acknowledgments}

This work is supported by Russian Foundation for Basic Research grant
14-02-01220 and by the Russian Ministry of Education and Science
contract 14.A12.31.0006.

\end{document}